\documentclass[12pt, a4paper]{iopart}

\usepackage{color}
\usepackage{graphicx}
\usepackage{float}
\usepackage{epsfig}
\usepackage{epstopdf}
\usepackage{subfig}
\usepackage{bm} 

\def\bt{{\bm t}}
\def\bu{{\bm u}}
\def\bw{{\bm w}}
\def\bB{{\bm B}}
\def\bF{{\bm F}}
\def\bJ{{\bm J}}
\def\brho{{\bm \rho}}

\def\grad{\nabla}
\def\curl{\nabla\times}
\def\div{\nabla\cdot}
\def\stressT{\stackrel{\leftrightarrow}{\bm{\pi}}}
\def\fsAver#1{\left\langle #1 \right\rangle} 

\def\begeqn{\begin{equation}}
\def\endeqn{\end{equation}}
\def\begeqnar{\begin{eqnarray}}
\def\endeqnar{\end{eqnarray}}
\def\begeqnarn{\begin{eqnarray*}}
\def\endeqnarn{\end{eqnarray*}}

\begin{document}

\title{Role of poloidal-pressure-asymmetry-driven flows in L-H transition and impurity transport during MGI shutdowns}
\author{A. Y. Aydemir$^1$, B. H. Park$^1$ and K. S. Han$^2$}
\address{$^1$National Fusion Research Institute, Daejeon 34133, Korea\\
$^2$University of Science and Technology, Daejeon, Korea}                        
\eads{aydemir@nfri.re.kr}

\section*{Abstract}
Poloidal asymmetries in tokamaks are usually investigated in the context of various transport processes, usually invoking neoclassical physics. A simpler approach based on magnetohydrodynamics (MHD), focusing on the effects rather than the causes of asymmetries, yields useful insights into the generation of shear flows and radial electric field.
The crucial point to recognize is that an MHD equilibrium in which the plasma pressure is not a flux function can be maintained only by contributions from mass flows.
Coupling between the asymmetry-generated forces and toroidal geometry results in a strongly up-down asymmetric effect, where the flows exhibit a strong dependence on the location of the asymmetry with respect to the midplane. This location-dependence can be used as an effective control mechanism for the edge and thus the global confinement in tokamaks. It can also explain a number of poorly-understood observations. For instance, strong dependence of the low to high (L-H) confinement transition power threshold $P_{LH}$ on the magnetic topology can be qualitatively explained  within this framework. Similarly, upper-lower midplane dependence of the poloidal flow direction after massive gas injections (MGI) naturally follows from this discussion. Similar arguments suggest that the ITER fueling ports above the midplane, to the extent  they can generate a positive pressure asymmetry at the edge, are misplaced and may lead to higher input power requirements.

\maketitle
\section{Introduction}

In an idealized magnetohydrodynamic (MHD) description of the axisymmetric plasma equilibrium, rapid parallel transport removes any density and temperature gradients along the field lines and leads to a highly symmetric state in which the plasma pressure is constant in both the poloidal and toroidal directions on flux surfaces.
In modern, diverted tokamaks, however, at least the poloidal symmetry is broken near the plasma edge due to changes in magnetic topology and the presence of strongly asymmetric perpendicular transport processes. Some commonly recognized sources of symmetry breaking are intrinsic error fields and externally imposed magnetic perturbations (see, for example, \cite{evans2015, in2017}). Possibly more importantly, the presence of a divertor in modern tokamaks and the accompanying neutral recycling near the X-point, a major source of fueling\cite{fukuda2000, groth2011},  results in a poloidally localized density and pressure increase within the separatrix in the same area\cite{carreras1998}. 
In addition, gas or pellet injection, either for fueling or as a disruption mitigation mechanism, are also fundamental sources of at least poloidal symmetry breaking (see Baylor\cite{baylor2007b} for ITER ports).

The potential for asymmetries to drive plasma flows have been recognized for quite sometime.
The first neoclassical studies of the role of poloidally asymmetric transport in the spin-up of a tokamak plasma can be traced back to Stringer\cite{stringer1969}. Others have expanded upon the ``Stringer mechanism'' to offer an explanation for the origin of the low to high (L-H) confinement transition\cite{wagner1982} in terms of the flows generated by inboard-outboard asymmetry of the neoclassical and turbulent particle fluxes\cite{hassam1991, hassam1993}. There have been also a number of studies examining the role of asymmetric neutral particle transport at the edge in driving toroidal flows (see, for example, Singh {\em et al.}\cite{singh2004}). However, in all these the emphasis has been on the asymmetry of the transport rather than the plasma profiles, and the theoretical treatment has involved intricate neoclassical arguments.
 
In his work we assume that the nonuniform distribution of particle sources at the plasma edge can lead to poloidally nonuniform density and pressure profiles. We show that the MHD equilibria consistent with these asymmetries necessarily have strong edge flows and driven radial electric fields. In turn, the flows and fields can have a profound effect on confinement through reduced turbulence\cite{shaing1989, hahm1995} and improved macroscopic stability\cite{bondeson1994, fitzpatrick1996, pustovitov2007b}. However, the focus of this work is on the generation of shear flows and radial electric field, not on their specific role in stability or confinement.
 
There is also a known inverse relationship between plasma dynamics and poloidal asymmetries. Centrifugal forces due to strong toroidal mass flows can introduce an in-out asymmetry in the density and pressure profiles\cite{zehrfeld1972, hameiri1983, guazzotto2004}.  Similarly, Alfv\'enic poloidal flows can lead to shocks with sharp poloidal density gradients\cite{shaing1992, seol2016}. Thus, there is a close connection between poloidal density or pressure asymmetries and mass flows. A great deal of this connection can be understood within a generalized MHD equilibrium framework. An earlier work\cite{aydemir2018c} addressed limited-tokamak equilibria. This work will focus on the more realistic, diverted configurations where the poloidal field at the edge is allowed to have one or more null points.

At this point a brief review of the axisymmetric tokamak equilibrium with flows will be useful.

\subsection{Some properties equilibrium flows}

Most general form of the plasma flow consistent with an axisymmetric MHD equilibrium can be obtained simply as follows\cite{zehrfeld1972, hameiri1983}: In steady-state, Faraday's law combined with the ideal MHD Ohm's law leads to $\bu\times\bB =\grad\phi,$ where $\phi$ is the electrostatic potential. Writing the axisymmetric field in the form $\bB=\grad\zeta\times\grad\psi + F\grad\zeta$, where $F=R^2\grad\zeta\cdot\bB$ and $\zeta$ is the usual toroidal angle, the resulting $E\times B$ velocity becomes
\begeqn
\bu_\perp  =   -\frac{\Omega F}{B^2}\bB + \Omega R^2\grad\zeta. \label{eqn:uPerp}
\endeqn
In general, the mass flux due to $\bu_\perp$ will not satisfy the steady-state continuity equation: $\div\rho_m\bu_\perp\ne 0$, where $\rho_m$ is the mass density. Thus, a parallel ``return flow'' (similar to the ``return'' Pfirsch-Schl\"uter current) is required to make the mass flux incompressible. Letting $\bu = (u_\parallel/B)\bB + \bu_\perp$ and requiring $\div\rho_m\bu = 0$ leads to
\begeqn
\frac{u_\parallel}{B} - \frac{\Omega F}{B^2}  =  \frac{\Phi(\psi)}{\rho_m},
\endeqn
where $\Phi$ is an arbitrary flux function. Combining with Eq.~\ref{eqn:uPerp} finally gives the general expression for the equilibrium flow velocity,
\begeqn
\bu = \frac{\Phi(\psi)}{\rho_m}\bB + \Omega(\psi) R^2\grad\zeta. \label{eqn:equilFlow}
\endeqn
A purely toroidal flow ($\Phi\rightarrow 0$) is possible only if there is a parallel flow of magnitude $u_\parallel = \Omega F/B$, which allows the poloidal projection of the parallel flow to cancel the poloidal component of $\bu_\perp.$ However, a purely poloidal flow requires an unrealistic density profile of the form $\rho_m =f(\psi)/R^2$, where $f=-\Phi F/\Omega$ is an arbitrary flux function.

In the presence of equilibrium flows, it is convenient to write the momentum equation in the form
\begeqn
\rho_m\frac{\partial\bu}{\partial t} = -\rho_m\grad(u^2/2) - \rho_m\bw\times\bu + \bJ\times\bB - \grad p  - \div\stressT,\label{eqn:momentum}
\endeqn
where $\bw \equiv \curl\bu$ is the vorticity, and $\stressT$ is the viscous stress tensor. In this work, the viscous stress tensor term is assumed to have the following form that is common in fluid calculations:
\begeqn
\div\stressT = \rho_m\nu \curl\curl \bu - \frac{4\rho_m\nu}{3}\grad\div\bu - \rho_m\gamma_p\bu_p, \label{eqn:stressT2}
\endeqn
where $\nu$ is the kinematic viscosity (momentum diffusivity), $\nu = \mu/\rho_m$, and $\mu$ is the usual scalar viscosity coefficient. The last term represents poloidal flow damping due to ``magnetic pumping''\cite{hassam1978}, where we assume $\gamma_p=0.68\nu_{ii}/\epsilon$\cite{shaing2015}. 

Here the scalar pressure term has to be treated carefully. The often-assumed adiabatic equation of state, with non-vanishing poloidal flows ($\Phi\ne 0$) leads to
\begeqn
\frac{p}{\rho_m^\gamma} = S(\psi), \label{eqn:adiabatic}
\endeqn
where $\gamma$ is the ratio of specific heats, and $S$, a measure of the entropy, is a flux function. However, in modern tokamaks, rapid parallel thermal transport ensures that the temperature itself is a flux function, $T=T(\psi)$, which is consistent with $p=\rho_mT$ and Eq.~\ref{eqn:adiabatic} only if $\gamma=1$ and $S\rightarrow T,$ thus effectively forcing an isothermal equation of state.

Assuming $\div\stressT=0$ for the moment and using an isothermal equation of state with $p=\rho_mT(\psi),$ the parallel component of Eq.~\ref{eqn:momentum} in steady-state leads to the ``Bernoulli equation''\cite{hameiri1983}
\begeqn
 \frac{\Phi^2B^2}{2\rho_m^2} + T\ln\rho_m - \frac{R^2\Omega^2}{2} = H(\psi), \label{eqn:bernoulli}
 \endeqn
where $H(\psi)$ is another arbitrary flux function. This equation, without the viscous stress tensor contribution, puts a constraint on the possible location of poloidal asymmetries. Taking the poloidal derivative gives
\begeqnar
\left(\frac{T(\psi)}{\rho_m} -  \frac{\Phi^2B^2}{\rho_m^3}\right) \frac{\partial\rho_m}{\partial\theta} & = &   \frac{\Omega^2}{2}\frac{\partial R^2}{\partial\theta} - \frac{\Phi^2}{2\rho_m^2}\frac{\partial B^2}{\partial\theta}. \label{eqn:constraint}
\endeqnar
For a tokamak we generally have $B\propto 1/R,$ and the two terms on the right-hand side have opposite signs since $(\partial R^2/\partial\theta) / (\partial B^2/\partial\theta) < 0.$ Thus, the density (and pressure) can have an extremum at a given point only if both terms vanish independently, which can happen only at the midplane, $\theta=\{0,\pi\}.$
A necessary conclusion is that to support a poloidal density extremum at an arbitrary point $\theta\ne \{0,\pi\}$,  the viscous stress tensor in Eq.~\ref{eqn:momentum} is needed. 

Although the arguments of this section are useful for an intuitive understanding of the basic properties of equilibrium flows, they omit the effects of particle sources. Also the discussion leading up to Eq.~\ref{eqn:equilFlow} assumes the existence of flux surfaces and fails near $X$-points. Effects of pressure asymmetries associated with external sources are introduced below. Complications due to poloidal field nulls have to be dealt with numerically and are discussed in subsequent sections.

\subsection{Poloidal torque and equilibrium calculations}

\begin{figure}[htbp]
\begin{center}
\includegraphics[height=2in]{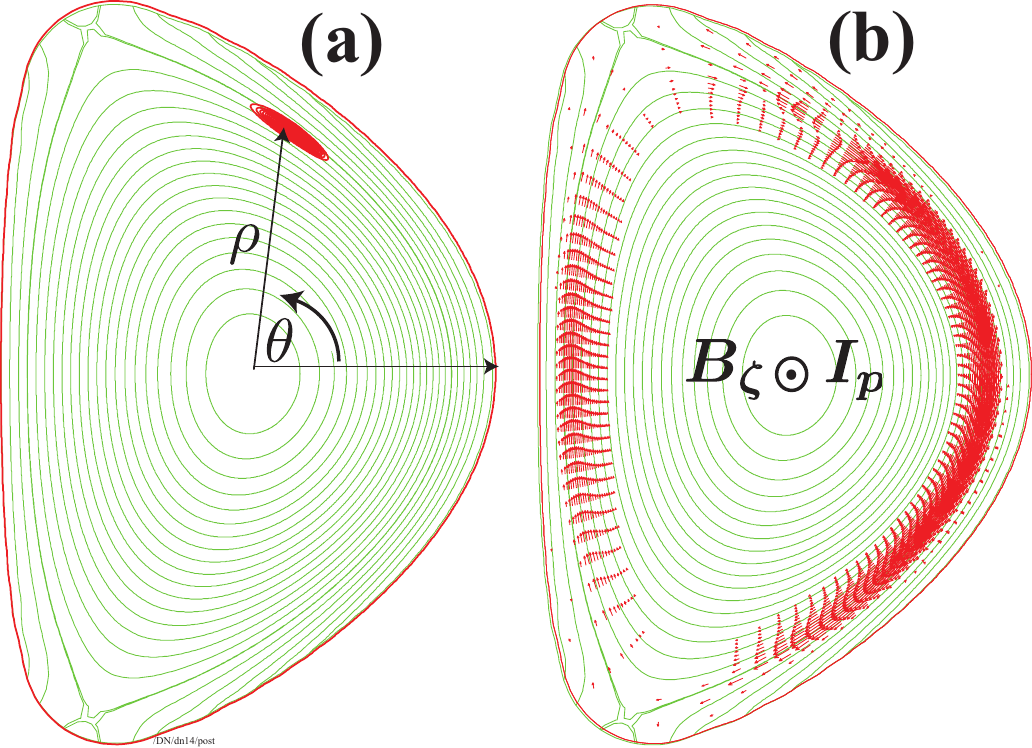}
\caption{\em \baselineskip 14pt (a) A localized positive pressure perturbation (red) near the separatrix in a double-null (DN) magnetic geometry. The wrapped Gaussian center is at $\theta_0=1.60$. (b)  Resulting negative poloidal shear flow. The toroidal field and plasma current are directed out of the plane of the figure (``the standard configuration''). The figures can be magnified arbitrarily to see the details.}
\label{fig:fig1_aydemir}
\end{center}
\end{figure}

Independent of any transport mechanism, a poloidal pressure asymmetry on a flux surface can generate a net torque resulting in poloidal and toroidal flows\cite{aydemir2018c}. This purely geometric effect can be understood easily: For convenience using axisymmetric flux coordinates $(\psi,\theta,\zeta)$ and starting with a localized pressure perturbation so that
\begeqn
p(\psi,\theta) = p_0(\psi) + \delta p(\psi,\theta), \label{eqn:p}
\endeqn
the poloidal force  $\bF_t$ within a flux surface and the torque $T_\zeta$ that drives a poloidal flow can be calculated as (see Fig.~\ref{fig:fig1_aydemir}(a))
\begeqnar
\bF_t & = & -\frac{1}{h_\theta^2}\bt(\bt\cdot\grad p),~~\bt \equiv \frac{\partial \brho}{\partial\theta},~~h_\theta \equiv |\bt|, \label{eqn:Ft} \\
T_\zeta & = & (\brho\times \bF_t)_\zeta. \label{eqn:Tzeta}
\endeqnar
For convenience we let $\delta p(\psi,\theta) = \delta p(\psi)f(\theta)$ and assume a wrapped Gaussian profile for $f(\theta)$ centered around $\theta=\theta_0$ to ensure periodicity. Then the net torque can be obtained by a simple surface average: $\fsAver{T_\zeta}_{s}=\oint T_\zeta dS_\psi/\oint dS_\psi.$ In circular geometry, it is given by\cite{aydemir2018c}

\begeqn
\fsAver{T_\zeta}_s  =  -\frac{r\delta p(r)}{2\pi R_0}\sum_{k=-\infty}^\infty  \oint e^{-(\theta-\theta_0 + 2\pi k)^2/w^2} \sin\theta d\theta. \label{eqn:averTorque}
\endeqn
For $w \ll 2\pi$, $\fsAver{T_\zeta}_s$ is approximately a sinusoidal function of $\theta_0$. The torque is positive if a positive pressure perturbation is in the lower half-plane $(\pi < \theta_0 < 2\pi)$, and vice versa. Note that the sign of the torque is entirely due to geometric effects in a torus and is independent of the current or toroidal field directions. 
But of course the radial electric field associated with the driven flows will be a function of the magnetic field.

Equation~\ref{eqn:averTorque} implies that the sign of the pressure perturbation, in addition to its location, plays an essential role; thus, it will be helpful to consider some experimental influences that broadly help determine the sign of $\delta p$. Neutral recycling near the $X$-point, and gas puffing or pellet injection, to the extent they add particles but not energy to the plasma, can be treated as being adiabatic to a first approximation. Therefore, for these processes we should have $\delta p/p \simeq \gamma\delta n/n,$ where $\gamma$ is the adiabatic index and $\delta n$ is the change in the particle density. Thus, naturally-occurring fueling through neutral recycling in the divertor chamber, or direct fueling using main-ion gas or pellets should lead to a positive pressure perturbation in quasi-steady state.
However, disruption mitigation efforts using massive gas or (shattered) pellet injection of high-Z materials (MGI, or  SPI) should generate a negative pressure perturbation since their goal
is the fast radiative collapse of the temperature (from keV to eV-range), a process that is by no means adiabatic. Thus, despite the increase in the particle density, we should expect $\delta p < 0$ for both MGI and SPI. Implications of this point will be discussed further in the section on massive impurity-injection-driven flows.
 
The equilibrium flows driven by the torque $T_\zeta$ of Eq.~\ref{eqn:averTorque} need to be calculated taking into account both the resulting asymmetric electromagnetic forces and various damping mechanisms. This is accomplished using the CTD code\cite{aydemir2015}, which solves the momentum equation shown in Eq~\ref{eqn:momentum}, with the stress tensor term shown in Eq.~\ref{eqn:stressT2}.
After perturbing a static, symmetric equilibrium with a perturbation of the form shown in Eq.~\ref{eqn:p}, a new equilibrium with flows is obtained in the asymptotic limit $t\rightarrow \infty,~\partial\bu/\partial t \rightarrow 0.$  During this relaxation process, the plasma current and toroidal flux are held constant using appropriate boundary conditions, while $\delta p$ is maintained externally.

Since configurations without a field null were discussed in some detail  earlier\cite{aydemir2018c}, here we will concentrate on geometries with one or more $X$-points. In an up-down asymmetric magnetic geometry, i.e., an upper or lower single null (USN, or LSN), various transport processes can generate average mass flows also\cite{strauss1995, aydemir2007a, aydemir2007b}. Therefore, the effects of  poloidal pressure asymmetries are best studied in isolation in a balanced double-null (DN) geometry, and that will be the focus of the next section.

\section{Poloidal pressure asymmetries in double-null (DN) geometry}
In a DN geometry,  transport-driven effects from the upper and lower nulls cancel, and without a pressure asymmetry there is no net flow or a radial electric field  within our MHD model; thus, $\fsAver{u_\theta}=\fsAver{u_\zeta}=\fsAver{E_\rho}=0.$ 
(From hereon $\fsAver{...}$ refers to the usual flux-surface average and not the ``surface-average'' of Eq.~\ref{eqn:averTorque}.)
With a pressure asymmetry above the midplane (Fig.~\ref{fig:fig1_aydemir}(a)), however, there is a negative average net torque, which in turn drives negative (in the ion diamagnetic drift direction) poloidal flows (Fig.~\ref{fig:fig1_aydemir}(b)). The shear flows are localized mostly around the separatrix. 

\begin{figure}[htbp]
\begin{center}
\includegraphics[height=2in]{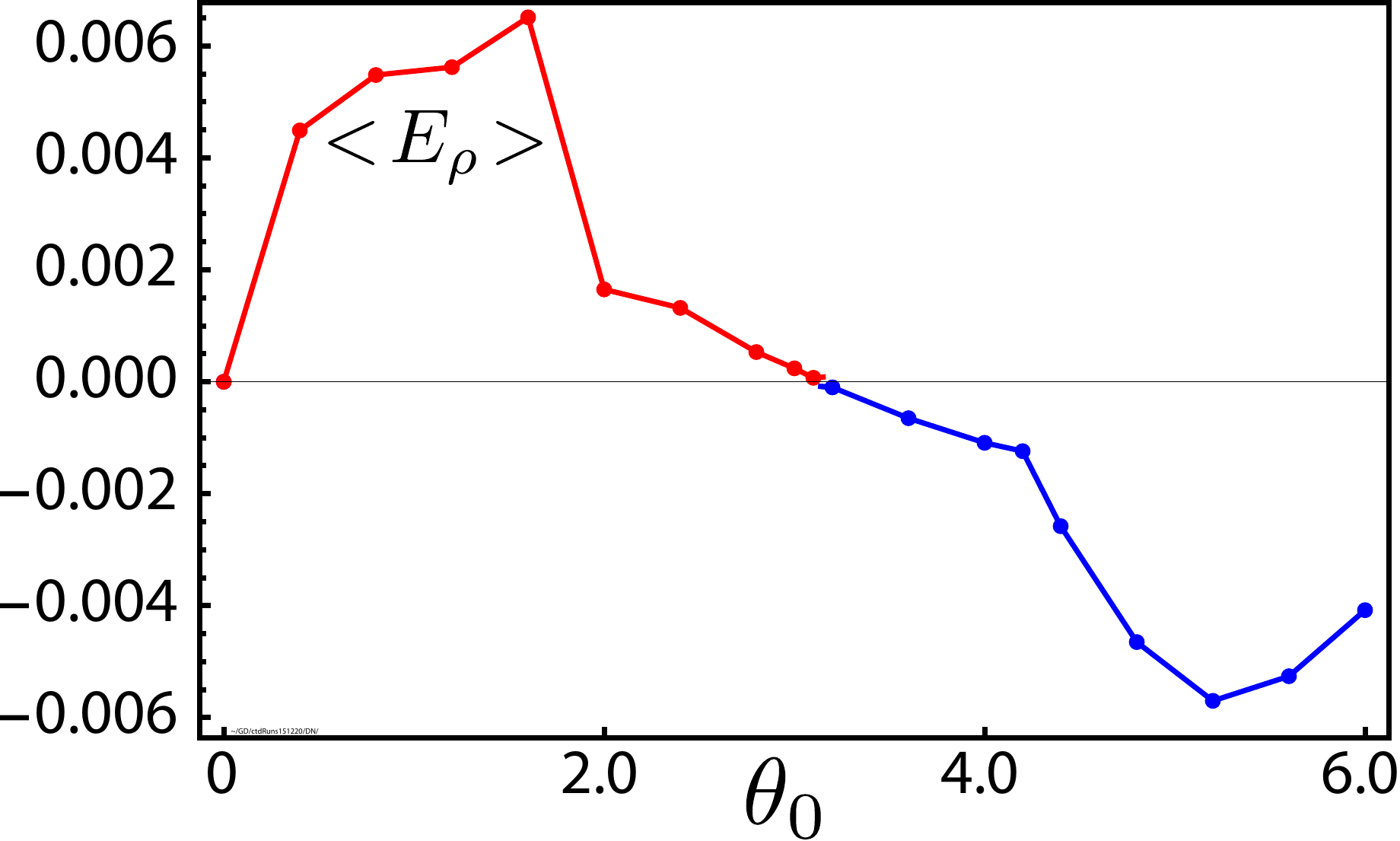}
\caption{\em \baselineskip 14pt Extremum of the flux-surface-averaged radial electric field as a function of the angle $\theta_0$, center of the wrapped Gaussian pressure perturbation.}
\label{fig:fig2_aydemir}
\end{center}
\end{figure}

\begin{figure}[htbp]
\begin{center}
\includegraphics[height=1.5in]{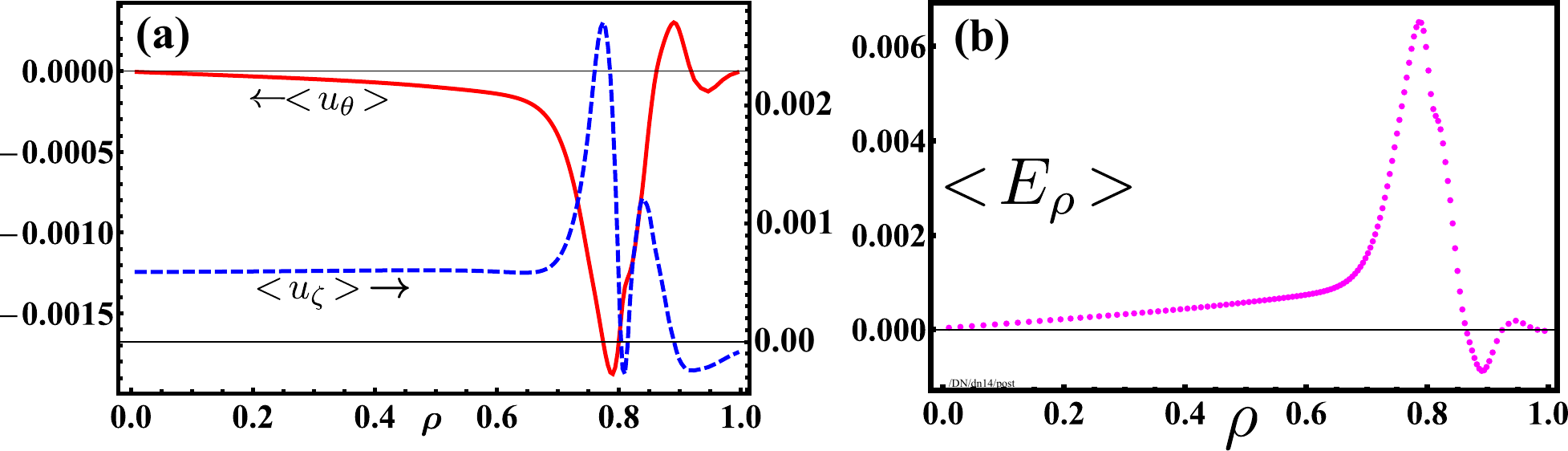}
\caption{\em \baselineskip 14pt (a) Flux-surface-averaged poloidal (solid red) and toroidal (dashed blue) velocities (normalized) produced by the pressure asymmetry of Fig.~\ref{fig:fig1_aydemir}(a) $(\theta_0=1.60)$. (b) Flux-surface-averaged radial electric field (normalized). The separatrix is at $\rho=0.81$.} 
\label{fig:fig3_aydemir}
\end{center}
\end{figure}

As suggested by the earlier discussion (and Eq.~\ref{eqn:averTorque}), location of the asymmetry has a strong influence on both the sign and amplitude of the shear flows, and the resulting radial electric field. This point is explicitly demonstrated in Fig.~\ref{fig:fig2_aydemir}, where the flux-surface-averaged field is plotted as a function of the poloidal angle $\theta_0$, center of the wrapped Gaussian used for the pressure perturbation (See Eq.~\ref{eqn:averTorque} and Fig.~\ref{fig:fig1_aydemir}). The figure shows the field extremum, which occurs near the separatrix (Fig.~\ref{fig:fig3_aydemir}(b)). If the points had been chosen more symmetrically about the midplane, we would expect the curve to exhibit more closely the simple sinusoidal symmetry $\fsAver{E_\rho}(\theta_0) =-\fsAver{E_\rho}(2\pi-\theta_0)$. It is clear, however, that the field extremum is positive for perturbations in the upper midplane ($0 < \theta_0 < \pi$) and negative for those in the lower midplane ($\pi < \theta_0 < 2\pi$), as implied by our earlier arguments.

Radial details of the flux-surface-averaged poloidal and toroidal velocities for the perturbation of Fig.~\ref{fig:fig1_aydemir} ($\theta_0=1.6$) are plotted in Fig.~\ref{fig:fig3_aydemir}(a), showing their highly sheared nature near the separatrix, which is located at $\rho=0.81$ as measured at the midplane. In Fig.~\ref{fig:fig3_aydemir}(b), the average radial electric field is shown, which exhibits a positive peak just inside the separatrix. Here the electric field is calculated using 
\begeqn
E_\rho = -u_\theta B_\zeta + u_\zeta B_\theta. \label{eqn:E}
\endeqn
Thus a large negative poloidal flow is correlated with a positive radial electric field.

In general, the shear flows and the associated electric field shown above would be expected to have a stabilizing influence on turbulence and the MHD modes localized near the separatrix. However, recall that other physics such as the ion orbit loss mechanism\cite{shaing1989} generate a {\em negative} radial electric field  inside the separatrix, and in general, the L-H transition is accompanied by a deepening electric-field well\cite{hahm1995}. Thus, the positive field in Fig.~\ref{fig:fig3_aydemir}(b) would oppose and possibly neutralize this process, and for this reason this location for the poloidal pressure asymmetry  would have to be considered as {\em unfavorable} from the point of view of turbulent transport and stability.

\begin{figure}[htbp]
\begin{center}
\includegraphics[height=1.5in]{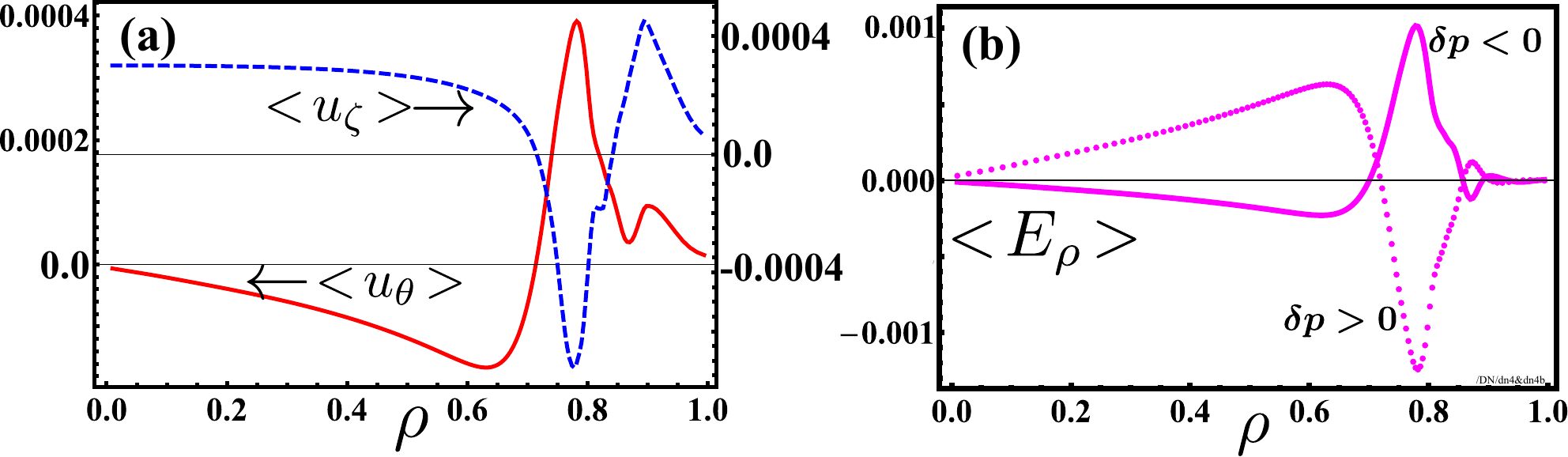}
\caption{\em \baselineskip 14pt (a) Flux-surface-averaged poloidal and toroidal velocities (normalized) produced by a positive pressure asymmetry at $\theta_0=4.2$, the lower X-point. (b) Flux-surface-averaged radial electric field for  $\delta p > 0$ (dotted), and $\delta p < 0$ (solid).}
\label{fig:fig4_aydemir}
\end{center}
\end{figure}

Moving the pressure asymmetry below the midplane, on the other hand, leads to quite favorable results. The deepest $E_\rho$ well is produced for $\theta_0\simeq 5.2$ (see Fig.~\ref{fig:fig2_aydemir}), an approximate mirror-reflection about the midplane of the perturbation in Fig.~\ref{fig:fig1_aydemir}(a). However, since we anticipate naturally-occurring pressure asymmetries near the $X$-points, we next examine a case with $\theta_0=4.2$, which places the perturbation near the lower $X$-point in Fig.~\ref{fig:fig1_aydemir}(a).
As seen in Fig.~\ref{fig:fig4_aydemir}(a) both the poloidal and toroidal shear flows reverse direction now, and a negative radial-electric-field well forms just inside the separatrix, with $E_\rho$ becoming positive further inside (Fig.~\ref{fig:fig4_aydemir}(b), the dotted curve labelled ``$\delta p > 0$''). This negative potential well should help with the L-H transition; in fact, it may even be sufficient to trigger it. Note that the positive toroidal rotation driven inside (where $\fsAver{E_\rho}>0$) is consistent with co-current ``intrinsic'' rotation observations after the L-H transition\cite{rice2004b} and also with its source being at the plasma edge.

In Fig.~\ref{fig:fig4_aydemir}(b) for $\fsAver{E_\rho}$, the curve labelled ``$\delta p < 0$'' is for a {\em negative} pressure perturbation of equal magnitude at the same location.  As the earlier simple torque-based arguments imply, with $\delta p < 0$ the torque reverses sign, leading to a reversal of the driven flows (not shown) and the resulting radial electric field (solid curve).

\section{Poloidal pressure asymmetries in single-null magnetic geometry}
In this section, the flows and radial electric field driven by pressure asymmetries in a lower-single-null (LSN) are investigated. Results for an upper-single-null (USN) are similar and can be obtained using symmetry arguments. As stated earlier, an up-down asymmetric magnetic geometry by itself can generate average mass flows\cite{strauss1995, aydemir2007a, aydemir2007b}; therefore, to isolate the effects of pressure asymmetry, a baseline LSN equilibrium with $\delta p=0$  that has only transport-driven flows is used for comparison. 

\begin{figure}[htbp]
\begin{center}
\includegraphics[height=1.5in]{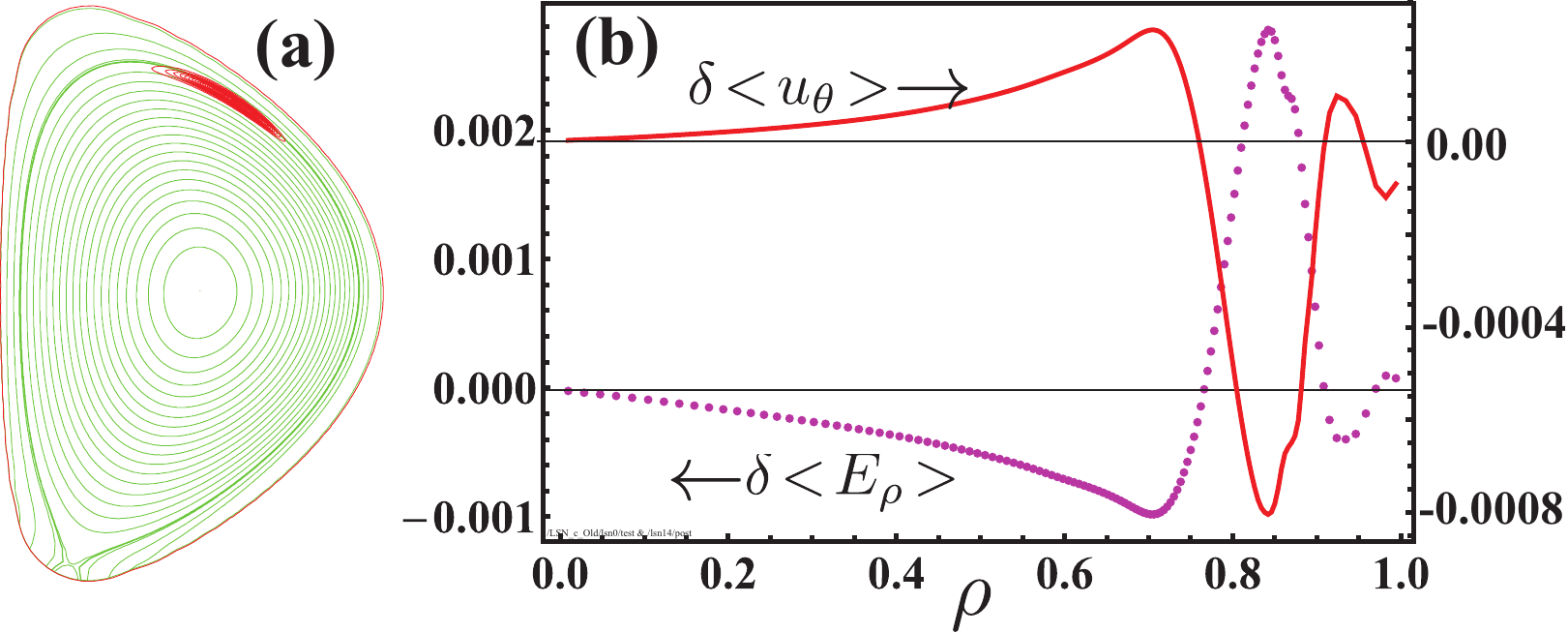}
\caption{\em \baselineskip 14pt LSN geometry. (a) Flux surfaces and a positive pressure perturbation at $\theta_0=1.6$, upper half-plane. (b) Flux-surface-averaged radial electric field (magenta, dotted) and the poloidal velocity (red, solid). Changes with respect to the baseline equilibrium with no perturbation are shown. The separatrix is at $\rho=0.86.$}
\label{fig:fig5_aydemir}
\end{center}
\end{figure}

Figure~\ref{fig:fig5_aydemir} corresponds to Fig.~\ref{fig:fig3_aydemir} of the DN configuration above and shows a positive perturbation above the midplane at $\theta_0=1.6$. Contribution to the electric field, now defined as
$\delta\fsAver{E_\rho}\equiv \fsAver{E_\rho} - \fsAver{E_\rho}_0$, where $\fsAver{E_\rho}_0$ is for the baseline equilibrium with $\delta p=0$, is again positive (Fig.~\ref{fig:fig5_aydemir}(b), driven by the negative poloidal flow near the separatrix.

\begin{figure}[htbp]
\begin{center}
\includegraphics[height=1.5in]{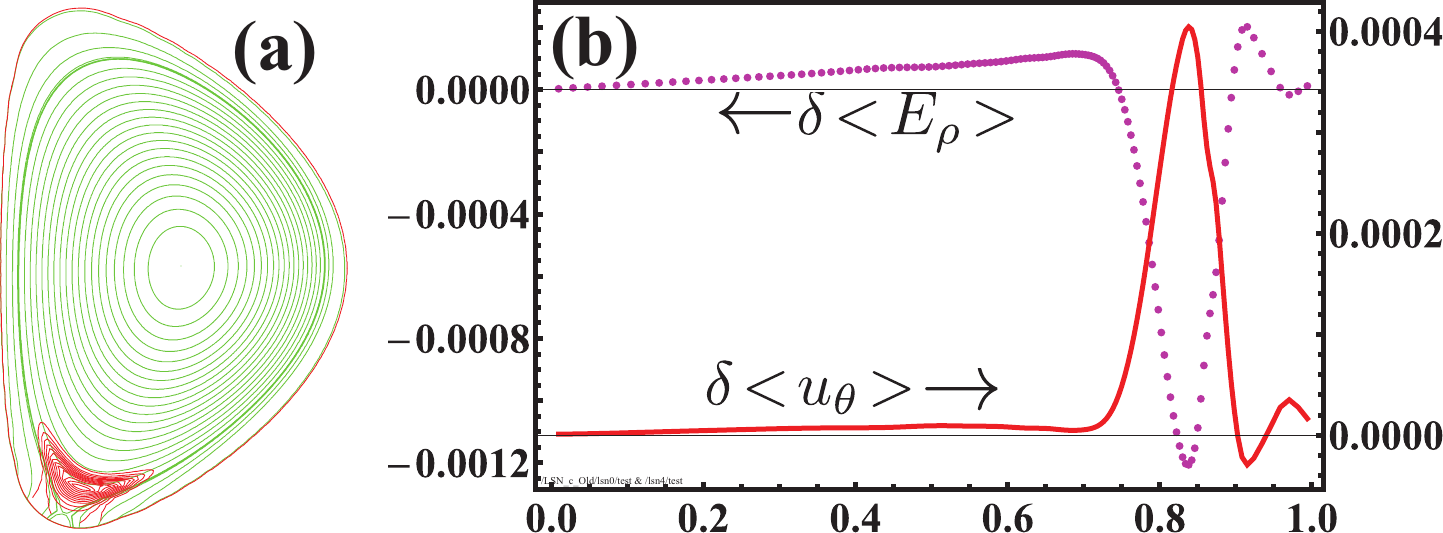}
\caption{\em \baselineskip 14pt LSN geometry. (a) Flux surfaces and a positive pressure perturbation at $\theta_0=4.2$, the X-point. (b) Flux-surface-averaged radial electric field (magenta, dotted) and the poloidal velocity (red, solid). Changes with respect to the baseline equilibrium with no perturbation are shown.}
\label{fig:fig6_aydemir}
\end{center}
\end{figure}

Similarly, Figure~\ref{fig:fig6_aydemir} corresponds to Fig.~\ref{fig:fig4_aydemir} of the DN configuration and shows a positive perturbation around the $X$-point. Because of the flux expansion near the null point, the perturbation in Fig.~\ref{fig:fig6_aydemir}(a) is wider in real space, although its width in $\psi$-space is the same as in Fig.~\ref{fig:fig5_aydemir}(a). Again, there is a pronounced and negative contribution to $\fsAver{E_\rho}$ by a positive poloidal flow localized around the separatrix (Fig.~\ref{fig:fig6_aydemir}(b)).

\section{Role in the L-H transition}

Recall that neutral recycling around the $X$-point (in the divertor chamber) is a significant source of plasma fueling\cite{groth2011, carreras1998}, which can lead to a localized, positive pressure perturbation in this region (this point may not be valid for the superD-X divertor geometry\cite{savarkar2018}); thus, we can expect the configuration of Fig.~\ref{fig:fig6_aydemir}(a) to occur naturally in diverted tokamaks. The resulting flows and radial electric field can play a significant role in the L-H transition, as already outlined in a preliminary version of this work\cite{aydemir2018c}. 

Unfortunately, despite its long history, we still lack a quantitative theory of how tokamaks spontaneously enter the H-mode; therefore, the arguments regarding the role of these poloidal asymmetries in the transition are necessarily qualitative, although they are robust and easily explained. Similar arguments were used earlier in discussions of transport-driven flows\cite{aydemir2012}. The crucial point here is that the L-H transition is accompanied by a deepening radial-electric-field well inside the separatrix (see, for example, \cite{burrell1990}). The primary driver of this field can be the ion orbit loss mechanism\cite{shaing1989}, or a combination of other physics not relevant to our discussion. We will assume that the electric field due to these processes is represented by $\fsAver{E_\rho}_{misc}$, with $\fsAver{E_\rho}_{misc} < 0$ at the edge. Then at any time the total electric field can be written as a sum of two terms:
\begeqn
\fsAver{E_\rho}_{tot} = \fsAver{E_\rho}_{misc} + \fsAver{E_\rho}_{\delta p}, \label{eqn:totE}
\endeqn
where $\fsAver{E_\rho}_{\delta p}$ is the contribution by a poloidally asymmetric pressure profile. Note that if we move beyond simple MHD and invoke, for example, a two-fluid theory, we would see that the radial ion pressure gradient also contributes to the electric field of Eq.~\ref{eqn:E}; we will assume this is contained within the $\fsAver{E_\rho}_{misc}$ term. The input power threshold for the transition, $P_{LH}$, is in general a complicated function of the plasma and machine parameters\cite{ryter1996}. Without making a causal connection, we can assume the transition is associated with a critical radial electric field level, $\fsAver{E_\rho}_{crit}$. Then we have $P_{LH} = f(\fsAver{E_\rho}_{crit})$, where  $f$ is a possibly machine-dependent function of the edge electric field.

\begin{figure}[htbp]
\begin{center}
\includegraphics[height=1.8in]{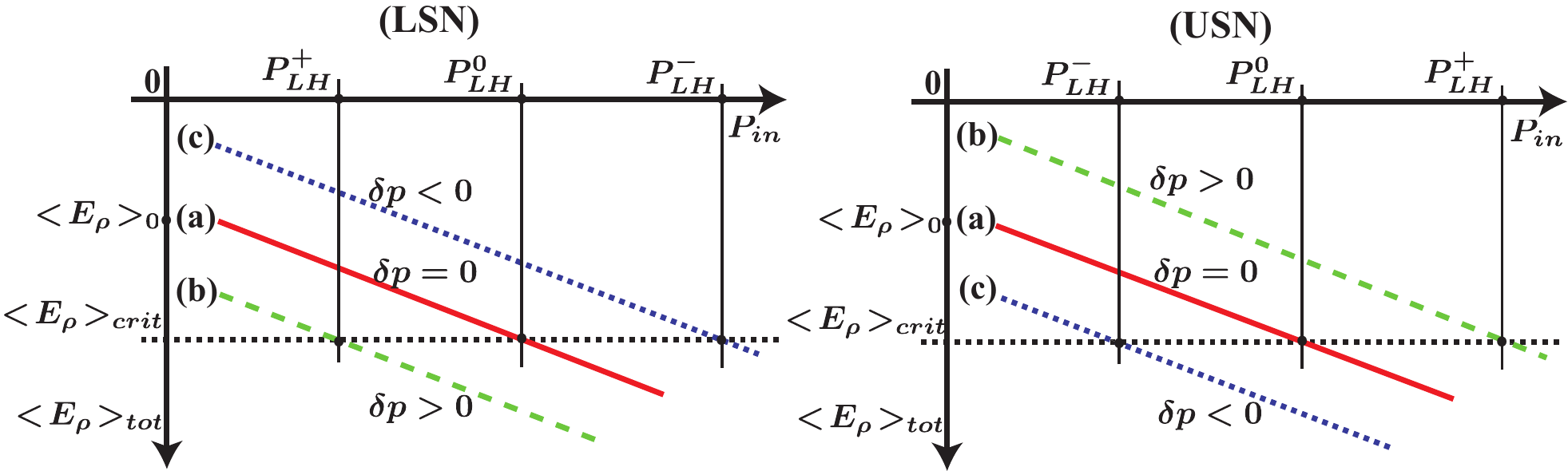}
\caption{\em \baselineskip 14pt Effect of a poloidal pressure asymmetry near the $X$-point on the L-H transition power threshold for lower and upper single-null magnetic geometries. ``Standard configuration'' of the fields is assumed.
(LSN): (a) No asymmetry, $\delta p = 0.$ (b) $\delta p > 0$, as in Fig.~\ref{fig:fig6_aydemir}(a). (c) $\delta p < 0$. Note that $P_{LH}^+ < P_{LH}^0$. (USN): Because the asymmetry is in the upper half-plane,  positions of the lines (b) and (c) are switched. Now $P_{LH}^+ > P_{LH}^0$.}
\label{fig:fig7_aydemir}
\end{center}
\end{figure}

Then in the input power versus edge electric field space shown in Fig.~\ref{fig:fig7_aydemir} we have the following picture, where we assume a linear relationship between $P_{in}$ and the edge electric field $\fsAver{E_\rho}_{tot}$ because we lack a quantitative theory of the L-H transition: 
\begin{itemize}
\item (LSN): Without a contribution from a poloidal asymmetry (line (a), $\delta p=0$), the radial electric field starts at some negative value, $\fsAver{E_\rho}_0$ and becomes progressively more negative as the input power is increased. The L-H transition is triggered as $\fsAver{E_\rho}_{tot}$ crosses the critical value at $\fsAver{E_\rho}_{crit},$ resulting in the power threshold $P_{LH}^0.$ Here changes in $\fsAver{E_\rho}_{tot}$ would be entirely due to $\fsAver{E_\rho}_{misc}$ of Eq.~\ref{eqn:totE}, driven by the increasing input power. If we allow for a poloidal pressure asymmetry near the $X$-point as in Fig.~\ref{fig:fig6_aydemir}(a), then the initial point is lower (line (b), $\delta p > 0$) due to the negative $\fsAver{E_\rho}_{\delta p}$ contribution by the asymmetry (see Fig.~\ref{fig:fig6_aydemir}) (b)). With a more negative starting point, less power is needed to cross the $\fsAver{E_\rho}_{crit}$ level, resulting in the lower value $P_{LH}^+ < P_{LH}^0$. 
A corollary of these arguments is that, if the asymmetry is negative as in line (c) (not very physical in the present context), then the initial point would be higher, since an asymmetry with $\delta p < 0$ in the lower half-plane leads to $\fsAver{E_\rho}_{\delta p} > 0$,  thus producing the higher value $P_{LH}^-$.

\item (USN): A positive pressure asymmetry near the $X$-point in USN magnetic geometry, because it is located in the upper half-plane, contributes a positive electric field, $\fsAver{E_\rho}_{\delta p} > 0$, thus reversing the role of the asymmetry in the L-H transition. Now the lines (b) and (c) switch positions, and we get $P_{LH}^+ > P_{LH}^0$.

\item As explained earlier, maintaining the LSN magnetic geometry but reversing the toroidal field reverses the sign of $\fsAver{E_\rho}_{\delta p}$ without reversing the sign of the poloidal flows (see also \cite{aydemir2007b}). Thus, the reversal $B_\zeta \rightarrow -B_\zeta$ would switch the positions of the lines (b) and (c) in the (LSN) panel of Fig.~\ref{fig:fig7_aydemir}, resulting in a state qualitatively like the (USN) panel. Now a positive pressure asymmetry would increase the threshold power, giving $P_{LH}^+ > P_{LH}^0$. 

\item The differences between the LSN and USN magnetic configurations in the standard configuration of the fields, and those between the standard and field-reversed states in the LSN configuration are consistent with the experimental observations where $P_{LH}$ increases approximately by a factor of two when the ion $\grad B$-drift direction points away from the active $X$-point\cite{ryter1996}. Note that in Fig.~\ref{fig:fig6_aydemir}(a) with the standard configuration of the fields, the drift is downward towards the $X$-point.

\item It is helpful to keep in mind that a negative perturbation in the lower half-plane produces poloidal flows in the same direction as a positive perturbation in the upper half-plane. This symmetry follows from the basic physics of the flow generation mechanism discussed earlier.
\end{itemize}

In a double-null (DN), effects of the positive pressure asymmetries at the two $X$-points would cancel each other, qualitatively leading to the state with $\delta p=0$ in the LSN or USN panels of Fig.~\ref{fig:fig7_aydemir}. Then we can conclude  that the expected power threshold levels for the three magnetic geometries, in the standard configuration of the fields, will have the order 
\begeqn
P_{LSN} < P_{DN} < P_{USN}. \label{eqn:P_LH_order}
\endeqn
Reversing the toroidal field direction will also reverse this order.

Up to now we were mainly concerned with positive pressure asymmetries produced by external fueling of the plasma using main ions. Next we examine poloidal flows due to negative pressure asymmetries produced by massive gas or pellet injections with a significant impurity content.

\section{Role in poloidal flows generated by impurity injections}
As mentioned earlier, impurities that accompany MGI or SPI for disruption mitigation necessarily leads to collapse of the plasma temperature from keV to eV range on a millisecond time scale. The result is a ``pressure hole'' near the injection site with $\delta p < 0$. The flows driven in this case are in opposite direction to those of the previous section where we had $\delta p > 0$ (analogous to the movement of an electron hole in a sea of electrons). And importantly, they can explain the poloidal flow patterns observed during the MGI shutdown events.

\begin{figure}[htbp]
\begin{center}
\includegraphics[width=6in]{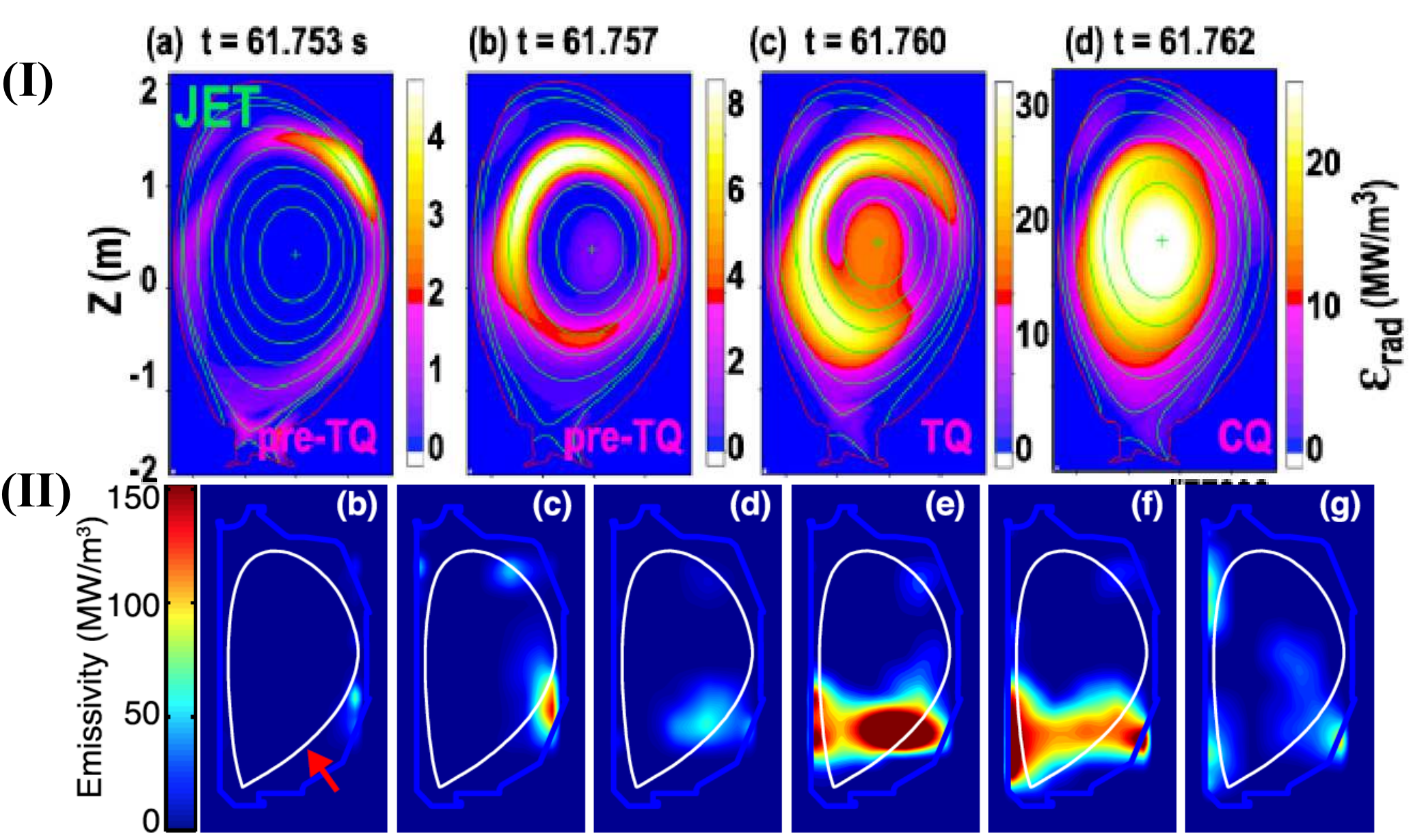}
\caption{\em \baselineskip 14pt (I) Counter-clockwise poloidal flows measured by fast bolometry during a MGI from an upper half-plane injection site in JET (reproduced with permission from Fig. 4 of \cite{hollmann2015}). (II) No flow or a possible clockwise flow during a MGI from a lower half-plane injection site in DIII-D (reproduced with permission from Fig.~3 of \cite{eidietis2017}). }
\label{fig:fig8_aydemir}
\end{center}
\end{figure}

The experimental observations are summarized in the two panels of Fig.~\ref{fig:fig8_aydemir}, reproduced here with permission from \cite{hollmann2015} and \cite{eidietis2017}. In (I), the top panel, ``[p]oloidal flows measured by
fast bolometry during MGI shutdowns [...] in JET'' are shown. Here the injection site is in the upper low-field side (LFS). Apparently the counter-clockwise flows, ``from the outer midplane over the top of the machine and down to the center post,'' shown in the figure are commonly seen in other devices also: see for example, Fig.~4 of \cite{hollmann2015} for DIII-D and AUG results, and Fig.~2 of \cite{eidietis2017} for more recent DIII-D observations.

When the injection site is in the lower half-plane, as in panel (II) of Fig.~\ref{fig:fig8_aydemir}, which is again from DIII-D (Fig.~3 of \cite{eidietis2017}), ``the emission exhibits either no poloidal flow and spreads directly across the plasma to the LFS or else a clockwise flow under the plasma and across the X-point''\cite{eidietis2017}. Moreover, both the upper and lower injection observations are independent of the toroidal field direction and thus have nothing to do with $E\times B$ flows\cite{eidietis2017},\cite{hollmann2013}.

However, the poloidal flows generated both by the upper and lower half-plane MGI injections,  and their insensitivity to the reversal of the toroidal field, can be explained in terms of the poloidal-pressure-asymmetry-driven flows. Although most of the physics has already been discussed, we will explicitly demonstrate the connection between MGI-generated negative pressure asymmetries and the experimental observations:

\begin{figure}[htbp]
\begin{center}
\includegraphics[width=6in]{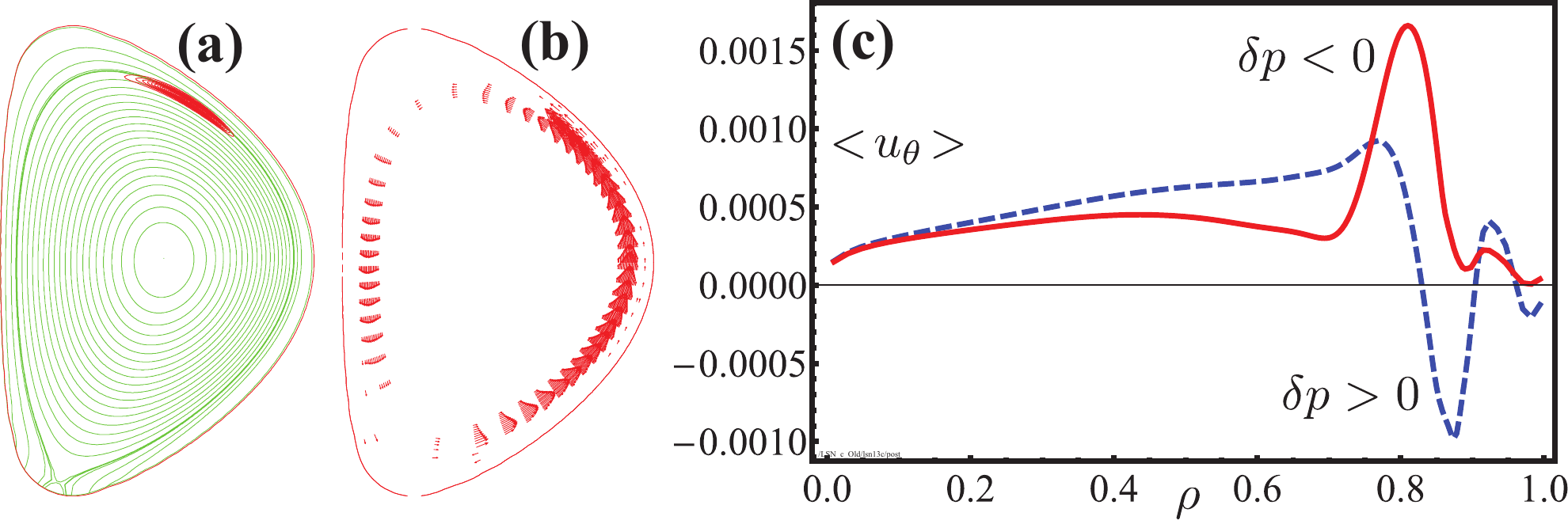}
\caption{\em \baselineskip 14pt Results of MGI in upper LFS. (a) Assumed negative pressure perturbation ($\delta p < 0$). (b) The resulting counter-clockwise poloidal flows. (c) Flux surface-averaged poloidal velocity for both $\delta p < 0$ (due to MGI, solid red line) and $\delta p > 0$ (due to fueling, dashed blue line). The separatrix is at $\rho=0.86$ at the midplane. The figure can be enlarged for a more detailed view. }
\label{fig:fig9_aydemir}
\end{center}
\end{figure}

\begin{figure}[htbp]
\begin{center}
\includegraphics[width=6in]{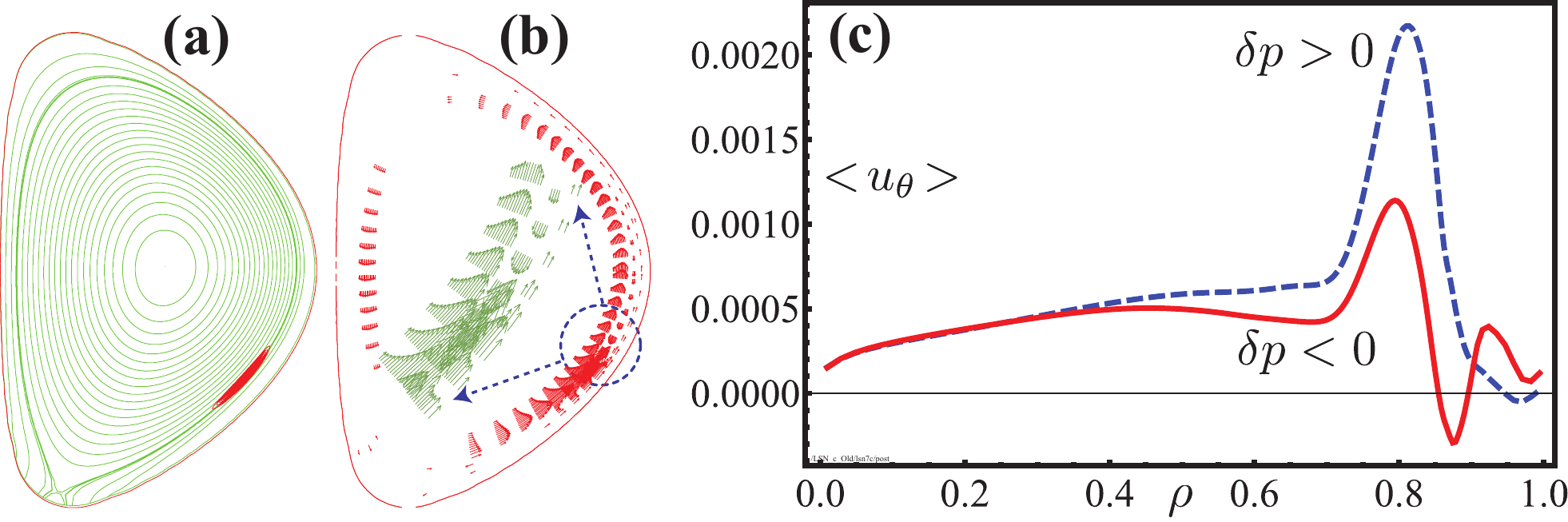}
\caption{\em \baselineskip 14pt Results of MGI in lower LFS. a) Location of the negative pressure perturbation ($\delta p < 0$). (b) The resulting poloidal flows with a stagnation point near the center of the MGI perturbation; the inset shows a blow-up of the stagnation area.
(c) Flux surface-averaged poloidal velocity for both $\delta p < 0$ (due to MGI, solid red line) and $\delta p > 0$ (due to fueling, dashed blue line).}
\label{fig:fig10_aydemir}
\end{center}
\end{figure}

\begin{itemize}
\item These flows are driven entirely by the toroidal geometry and their direction is determined by the location of the asymmetry with respect to the midplane. Asymmetries with $\delta p > 0$ in the upper half-plane produce negative (clockwise) flows. This was shown unambiguously in the DN magnetic geometry of Fig.~\ref{fig:fig1_aydemir}, where other sources of flows cancel.
\item The driving torque is independent of the fields (see Eq.~\ref{eqn:averTorque}); the resulting poloidal flows are independent of the toroidal field direction, although the sign of the generated radial electric field $E_\rho$ does depend on $B_\zeta$.
\item The driving torque, the resulting poloidal flows and radial electric field all reverse directions when $\delta p < 0$ (see Eq.~\ref{eqn:averTorque} and Fig.~\ref{fig:fig4_aydemir} (b)).
\item In LSN magnetic geometry, the expected positive pressure asymmetry around the $X$-point makes a positive contribution to the transport-driven poloidal flows, as we saw in Fig.~\ref{fig:fig6_aydemir}. An MGI-produced negative asymmetry in the upper half-plane also adds constructively and results in the positive (counter-clockwise) flows as seen Fig.~\ref{fig:fig9_aydemir}.
Panel (a) shows the assumed negative pressure perturbation ($\delta p < 0$). Panel (b) shows the resulting counter-clockwise poloidal flows, which would normally take the radiation pattern over the top to the HFS. Recall, however, that these are equilibrium calculations in which we seek a quasi steady-state with flows in the presence of a prescribed poloidal asymmetry that is held stationary. Thus, the effect of the flows on the pressure asymmetry is not calculated here and left for a future work.
For comparison, in (c) the flux surface-averaged poloidal velocity for both $\delta p < 0$ (due to MGI, solid red line) and $\delta p > 0$ (due to fueling, dashed blue line) are shown. Note that although fueling at this location would reverse the flows (see Fig.~\ref{fig:fig5_aydemir} (b)), MGI makes a strong positive contribution and the total flow shown here is positive.

\item  Results of MGI in lower LFS are quite different, as seen in Fig.~\ref{fig:fig10_aydemir}. Again the panel (a) shows the location of the negative pressure perturbation ($\delta p < 0$). The resulting poloidal flows are seen in (b). On the inner flux surfaces, the transport-driven flows continue nearly unchanged, but on the outer surfaces affected by MGI, the poloidal flow reverses, creating a stagnation point near the center of the MGI perturbation, where $u_\theta \simeq 0.$
Until the temperature collapses due to radiation and sets up $\delta p < 0$,  it is possible the existing positive flows may carry injected the material up towards the midplane. But once the pressure hole is established, it is clear the radiation pattern will be stationary or move downward with the negative flows, consistent with the experimental observations in Fig.~\ref{fig:fig8_aydemir}. As stated earlier, these equilibrium calculations do not follow the time evolution of the asymmetry; this interesting exercise is left for a future work.

\end{itemize}

\section{Discussion and Summary}
The numerical results above were presented in non-dimensional units. In order to get a sense of the possible magnitude of the actual flows and electric fields generated, we will use the following parameters for the edge plasma (similar to those used in \cite{aydemir2018c}): Toroidal field $B_{\zeta 0}=3T$, deuterium density $n=10^{19}m^{-3}$, minor radius $a=1m$,  and the inverse aspect ratio $\epsilon=a/R_0=1/3$. With these we get the poloidal Alfv\'en speed $v_{Ap} = \epsilon B_{\zeta 0}/\sqrt{\mu_0\rho_m}=5\times 10^6 ms^{-1}$. Thus the flow velocities, normalized with $v_{Ap}$, are of order $10^{-3}v_{Ap} = 5km s^{-1}$. The electric field is normalized with $E_0=\epsilon v_{Ap}B_{\zeta 0} = 5\times 10^6 Vm^{-1}$,
which leads to $\fsAver{E_\rho}_{min}\simeq 6\times 10^{-3}E_0= 30kVm^{-1}$ for Fig.~\ref{fig:fig2_aydemir},  a substantial radial electric field. With the same parameters we get $\fsAver{E_\rho}_{max}\simeq +10kVm^{-1}$ for Fig.~\ref{fig:fig5_aydemir} (b), clearly indicating that any density or pressure asymmetry above the low-field-side midplane (an ITER gas injection site) would be detrimental to confinement.

The quasi-steady state values for $\fsAver{u_\theta}$ and $\fsAver{E_\rho}$ are of course affected by the assumed transport coefficients. Typical values used for the poloidal damping rate $\gamma_p$ and viscosity coefficient $\mu=\rho_m\nu$ that appear in  Eq.~\ref{eqn:stressT2} are as follows: $\gamma_p=10^{-4}$, $\mu=5.0\times 10^{-6}$, or in dimensional units $\gamma_p=10^{-4}(v_{Ap}/a)\simeq 5\times 10^2 s^{-1}$ and $\mu=5\times 10^{-6}(av_{Ap})=25m^2s^{-1}$, both of which are somewhat larger than physical estimates.

Since the torque $\fsAver{T_\zeta}$ (but not necessarily the full MHD results) is linear in the perturbation amplitude $\delta p$ (see Eq.~\ref{eqn:averTorque}), and since this quantity is not easily available experimentally, our numerical results should be interpreted in the light of this uncertainty. The numerical calculations typically used $\delta p/p_0\sim 10^{-4}$, where $p_0$ is the pressure on axis.

In summary, a relatively small-amplitude poloidal pressure asymmetry at the plasma edge, possibly maintained by a neutral particle source, is shown to drive substantial mass flows near the plasma boundary, entirely within an MHD equilibrium framework. The mechanism is robust and relies essentially on force-balance arguments. Whereas the transport-theory based analyses of poloidal asymmetries tend to focus on inboard vs. outboard differences, we find that the position with respect to the midplane plays a more significant role. For the ``standard configuration'' of the plasma current and toroidal field, asymmetries below the midplane produce a negative radial electric field inside the separatrix. This field can then enhance or even supplant other sources of edge radial electric field (e.g.,  the ion orbit-loss), and thus play a significant role in suppressing edge turbulence and MHD activity. The result would be an easier L-H transition and a lower transition power threshold, $P_{LH}.$

The generated poloidal and toroidal flows, and the associated $E_\rho$ can be easily shown to have the correct symmetries\cite{aydemir2007b} to explain the ion $\grad B$-drift-dependence of $P_{LH}$. Briefly, the sign of the net torque and the poloidal flow due to the asymmetry depends only on its location with respect to the midplane. In a discharge in the ``standard configuration,'' if the asymmetry is near the lower $X$-point, for instance, due to neutral recycling,  reversing the toroidal field will leave the poloidal (but not the toroidal) velocity intact and thus reverse $E_\rho,$ along with the ion drift direction. These changes will convert the original favorable configuration into an unfavorable one with a higher $P_{LH},$ consistent with experimental observations.

Note that one of the ITER fueling ports\cite{baylor2007b} is approximately at the location of the asymmetry in Fig.~\ref{fig:fig1_aydemir}(a). If fueling at that location can penetrate the plasma edge and produce a poloidal pressure asymmetry, it will generate negative shear flows and a positive radial electric field, which will tend to increase the L-H transition power threshold. Fortunately other ports near the $X$-point are quite favorably located from this point of view.

More generally, since the edge is the most easily accessible part of the discharge, a deliberately-induced poloidal asymmetry at an appropriate location along the plasma periphery can be used as an effective tool in generating highly beneficial edge shear flows. These flows, depending on their location with respect to the midplane, can be used to enhance or degrade edge confinement as needed.

Finally, MGI-produced pressure holes ($\delta p < 0$) near the plasma boundary drive poloidal flows in opposite direction to those that result from $\delta p>0$. These can explain the quite different flow patterns seen after MGI at the upper and lower LFS injection sites in DIII-D and other devices.

\section*{Acknowledgements}
This work was supported by MSIP, the Korean Ministry of Science, ICT and Future Planning, through the KSTAR project.

\section*{References}

\end{document}